\documentclass[british]{elsarticle}
\usepackage[T1]{fontenc}
\usepackage[latin9]{inputenc}
\usepackage{parskip}
\usepackage{color}
\usepackage{url}
\usepackage{amstext}
\usepackage{graphicx}

\makeatletter

\newcommand{\lyxmathsym}[1]{\ifmmode\begingroup\def\b@ld{bold}
  \text{\ifx\math@version\b@ld\bfseries\fi#1}\endgroup\else#1\fi}

\makeatother

\usepackage{babel}
\begin{document}
\begin{frontmatter}
\title{The Application of VMM3A Readout for Multi-Grid Neutron Detectors}
\author[ess]{A. Backis}
\author[ess]{F. Piscitelli}
\author[ess]{D. Pfeiffer}
\author[gla]{J.R.M. Annand\label{cauth}}
\ead{john.annand@glasgow.ac.uk}
\author[ess]{M. Aouane}
\author[ess]{K. G. Fissum}
\author[gla]{K. Livingston}
\author[ISIS]{G. Mauri}
\author[ISIS]{D. Raspino}
\address[ess]{Detector Group, European Spallation Source ERIC, SE-221 00 Lund, Sweden}
\address[gla]{School of Physics and Astronomy, University of Glasgow G12 8QQ, Scotland,
UK}
\address[ISIS]{ISIS Facility, Rutherford Appleton Laboratory, Harwell Campus, Oxfordshire
OX11 0QX, UK}
\cortext[cauth]{Corresponding Author}
\begin{abstract}
The T-REX neutron spectrometer at the European Spallation Source will
use Multi-Grid Technology, which is a voxelised proportional counter
relying on$\mathrm{^{10}B_{4}C}$ coatings to detect the scattered
neutrons. Measurements of the position dependence of pulse-height
and relative detection efficiency of a Multi-Grid prototype of the
T-REX spectrometer are presented for two different schemes of signal-processing
electronics based on the VMM3A ASIC and CREMAT technology. These measurements,
intended to test the suitability of VMM3A for readout of the T-REX
Multi-Grid, are compared with Monte Carlo simulations based on the
Garfield++ and Geant4 tool kits.
\end{abstract}
\end{frontmatter}

\section{Introduction}

\begin{figure}[h]
\includegraphics[width=1\columnwidth]{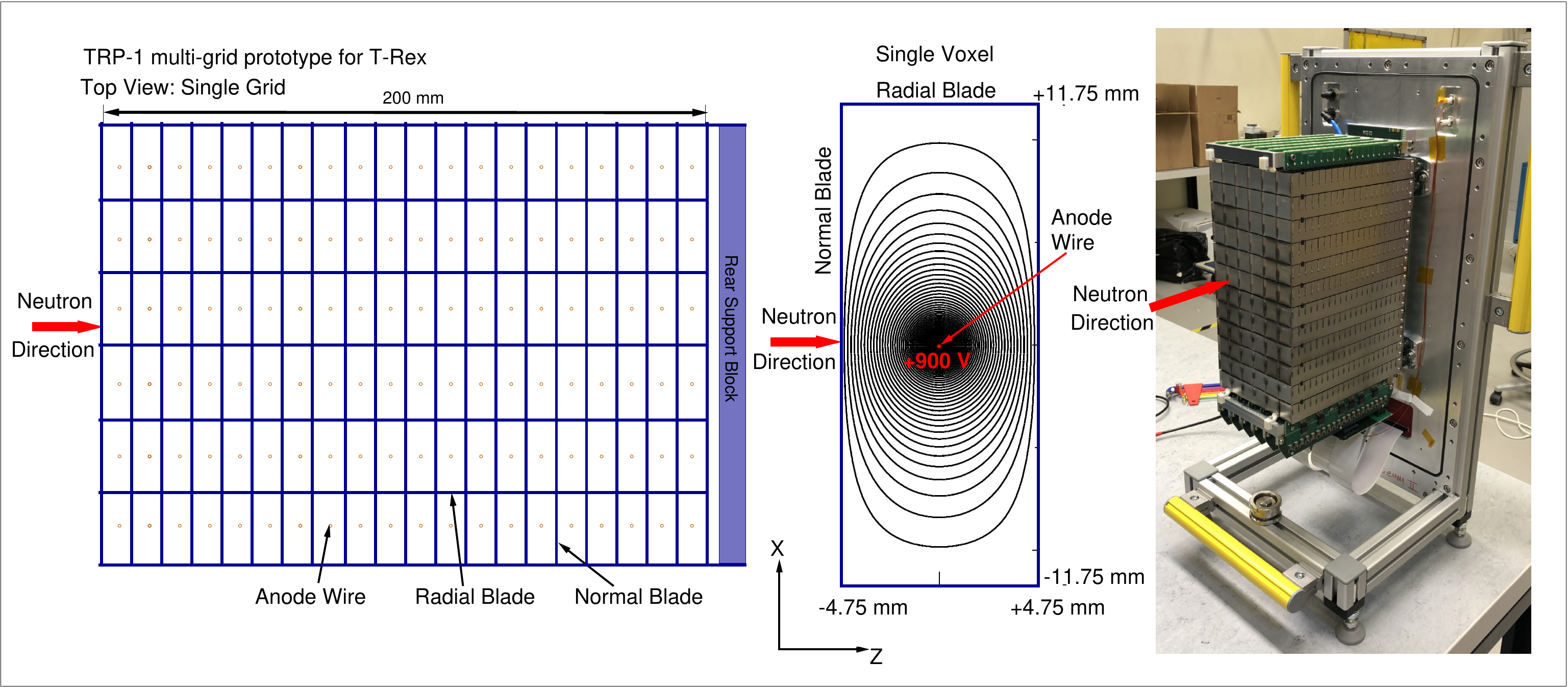}

\caption{\protect\label{fig:Multigrid}Left: a single grid from the T-REX prototype
TRP-1. Middle: a single voxel of the grid showing equipotential contours
calculated by Garfield++. Right: a photograph of TRP-1 at ESS with
the outer gas enclosure vessel removed, showing the stack of 12 grids.}
\end{figure}

Multi-Grid (MG) technology for detection of thermal or cold neutrons,
originally developed at ILL \citep{ILL}, will be used for the T-REX
bispectral chopper spectrometer to be installed at the European Spallation
Source (ESS) \citep{ESS}. MG is a vertical stack of grids, each a
rectangular lattice of normal and radial Al blades (Fig.\ref{fig:Multigrid}).
The grids form the cathodes of a voxelised proportional counter (VPC),
with wires strung vertically through the centres of each voxel providing
the anodes. The normal blades are coated at ESS with $\mathrm{^{10}B}$-enriched
(97\%) $\mathrm{B_{4}C}$ \citep{B4Cfilm}, and neutron capture in
this film of $\mathrm{B_{4}C}$ produces $\mathrm{^{4}He}$ and $\mathrm{^{7}Li}$
residual nuclei, one of which escapes the film into the VPC gas (
Ar-CO$_{2}$ in the present case) generating a signal. 

This work describes a comparison of signal properties for two alternative
schemes to read out the charge collected from the MG electrodes. This
was performed using the EMMA thermal neutron beam \citep{EMMA} at
the ISIS neutron spallation source in the UK.

\section{Technical Details}

The MG voxel, viewed from the top in Fig.~\ref{fig:Multigrid}, is
an elongated rectangle measuring $\mathrm{23.5\times9.5\:mm^{2}}$
in the $x-z$ plane, while it is $\mathrm{23.5\times24.0\:mm^{2}}$
in the $x-y$ plane. When a positive high voltage (HV) is applied
to the anode wire, the corners of the voxel experience very low voltage
gradients compared to more central regions. This affects the charge
collection time, which increases as the neutron-capture position moves
from the centre at $x=0.0$~mm to a corner at $x=\pm11.75$~mm.
Neutrons are captured on the normal-blade coatings at $z=\pm4.75$~mm.

T-REX prototype detector TRP-1 was tested. Unlike subsequent T-REX
prototypes, this has no internal neutron shielding, such as $\mathrm{B_{4}C}$
coating on the radial blades, but this does not affect charge collection.
The absence of capture events on radial blades makes the understanding
of the pulse-height response simpler. The coating thickness on the
normal blades is $\approx1\:\mu$m on the front voxels, rising to
$\approx2\:\mathrm{\mu m}$ on the rear voxels. As an important goal
here was to compare different electronic readout systems, the same
detector was connected to the two sets of electronics based on the
VMM3A ASIC and CREMAT technology described in Sec.~\ref{subsec:VMM3A}
and Sec.~\ref{subsec:CREMAT} respectively.

\subsection{\protect\label{subsec:Electronics}Electronics}

\begin{figure}[h]
\includegraphics[width=1\columnwidth]{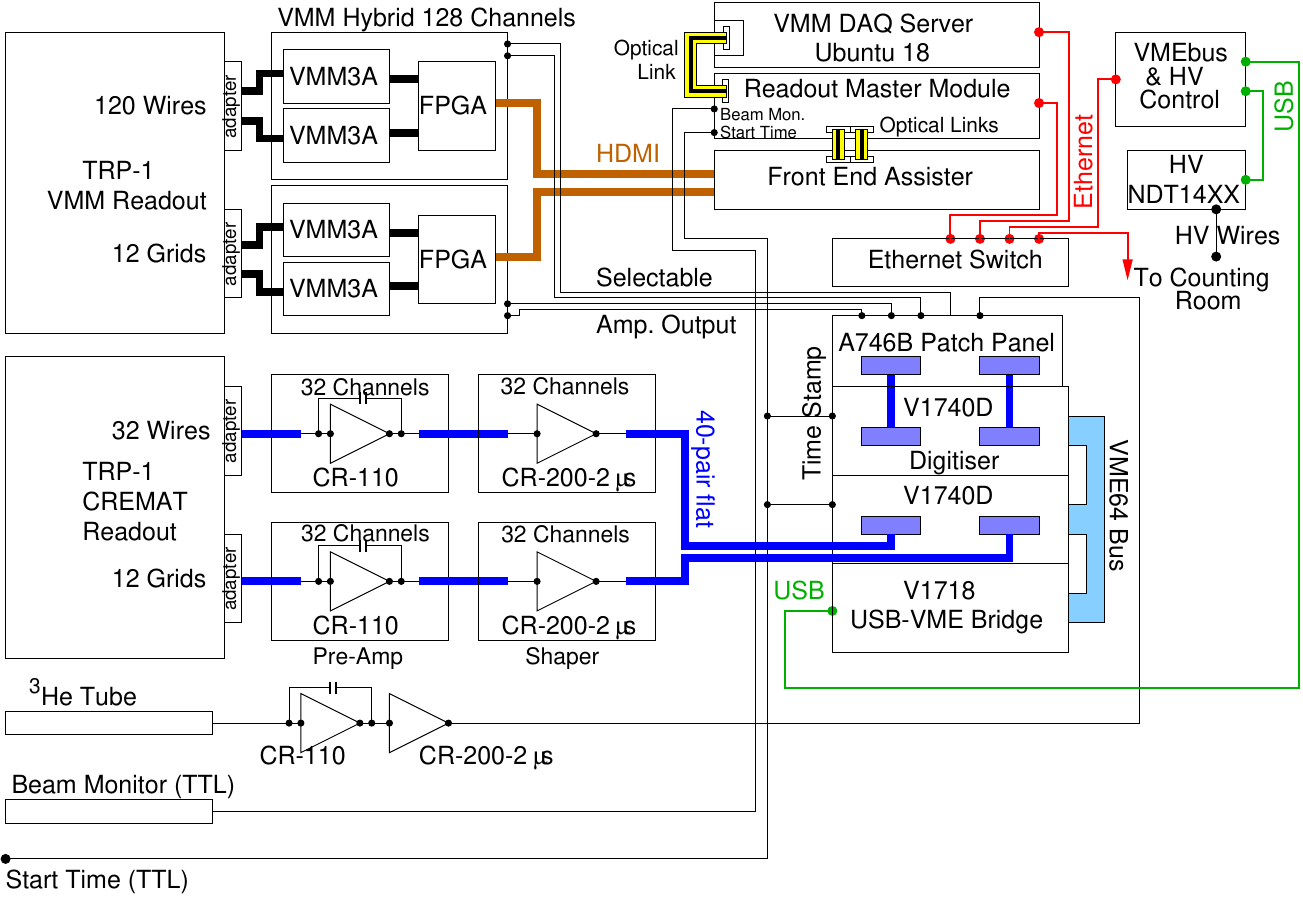}

\caption{\protect\label{fig:Electronic-layout}Schematic diagram of the readout
electronics. Either VMM hybrids or CREMAT pre-amps were connected
to the TRP-1 wires and grids via adapter cards. }
\end{figure}

The major motivation for this work was to test the suitability of
read-out electronics based on the VMM3A ASIC for T-REX. VMM3A is used
on the Multi-Blade detector \citep{Multiblade}, employed on three
instruments at ESS, and is available in sufficient quantity to instrument
T-REX. However, due to relatively slow charge collection in the MG,
the maximum pulse-shaping time constant on the front-end pulse amplifier
is shorter than optimum for a MG-type VPC. Thus CREMAT amplifiers
with longer shaping times were also tested for comparison. The latter
however are not suitable for instrumentation of T-REX, which will
contain 60, 88-grid columns of MG, giving a total of 12480 channels
to read out. A schematic of the readout electronics is displayed in
Fig.~\ref{fig:Electronic-layout}. TRP-1 has 12 grids (cathodes)
and 120 wires, giving 132 channels and 1440 voxels in total. Signal
lines attached to the wire and grid electrodes were fed through the
gas containment vessel of TRP-1 and connected to either VMM Hybrid
or CREMAT cards. With the VMM Hybrid all channels were recorded, while
for CREMAT 32 wires and 12 grids were recorded due to a limited supply
of this hardware. High voltage was supplied by a CAEN \textcolor{black}{NDT1470}
under software control.

In addition to TRP-1 the data acquisition system (DAQ) also accepted
the Start Time from the accelerator for time of flight (TOF) determination,
the EMMA beam monitor for neutron flux normalisation and optionally
a $\mathrm{^{3}He}$ tube attached to CREMAT electronics for comparison
of detection efficiency.

\subsubsection{\protect\label{subsec:VMM3A}VMM3A}

VMM3A \citep{VMM3A} is a 64-channel ASIC \citep{VMM3A,VMM3A_1},
developed at Brookhaven National Laboratory for ATLAS at CERN, and
designed to readout Micromegas and sTGC detectors, which are considerably
faster than a MG VPC. It is a high density system with fast data transfer
capability, suitable for large numbers of readout channels.

Each VMM3A channel contains an analogue section consisting of a charge
sensitive preamplifier, a shaping amplifier a discriminator and a
peak finder. The gain of the system can be adjusted over a range of
0.5 mV\ensuremath{\slash}fC to 16.0 mV\ensuremath{\slash}fC, and the
peaking time of the shaper can be set to values between 25 ns and
200 ns. Peak pulse height (PH) and trigger time are then digitised
on-chip. The pulse peaking time was set to the maximum of 200~ns,
which is shorter than optimum for MG detectors, while the gain was
set to 3 mV/fC for anode wires and 4.5 mV/fC for cathode grids. 

The VMM3A ASICs are integrated into the `Scalable Readout System'
\citep{VMM3A_1} developed by the RD51 collaboration. Two ASICs are
mounted on a VMM hybrid card, giving a total of 128 channels. The
hybrid also mounts a Spartan-7 FPGA which reads the digitised pulse
height, time stamp and channel ID from the ASICs and passes the assembled
data to the Front End Assistor (FEA) via HDMI cables. On this system
the FEA receives LVDS signals, converts to a serial optical standard
and sends the data to a Readout Master Module (RMM) via optical fibres.
\textcolor{black}{The RMM also receives the TTL start time and BM
output, which are incorporated in the data stream and passed to the
VMM DAQ Server via optical fibres.} Both the FEA and RMM have been
developed at ESS.

Low voltage power to the VMM hybrids is supplied by the FEA. The hybrids
also mount 00-Lemo ports to which selected signals from the VMM3A
amplifiers can be routed under software control. These signals were
sent to a CAEN V1740D digitiser \citep{V1470D}, which recorded the
waveform. Note that the VMM3A pulse-height spectra displayed in this
article are derived from the peak pulse height digitised on-chip.

\subsubsection{\protect\label{subsec:CREMAT}CREMAT}

Cremat amplifiers \citep{CREMAT} read out by a \textcolor{black}{CAEN
V1740D pu}lse digitiser \citep{V1470D} are quite flexible to use
and convenient for testing small prototypes, but the density of channels
to too low for instrumentation of large spectrometers. Charge from
the MG electrodes was collected by CR-110 charge-sensitive preamplifiers
coupled to CR-200-2$\mu$s shaper amplifiers. These were mounted on
32-channel PCBs produced at ESS. The pre-amp gain was 1.4 mV/fC, while
the shaper gain was 10 followed by a resistive attenuation factor
of 6, giving a final gain of 2.3 mV/fC. Attenuation was required to
bring the pulse amplitude within the dynamic range of the subsequen\textcolor{black}{t
pu}lse digitiser which operated in two modes: either recording the
waveform produced by the CREMAT amplifier, or integrating the waveform
over a time period of\textcolor{black}{{} 10~$\mu s$ }to produce the
pulse amplitude. Integration was performed by DPP-QDC firmware from
CAEN. Pulse amplitude, time stamp and channel ID were then read from
the digitisers by the VME \& HV control laptop via the VMEbus and
a CAEN V1718 VME-USB bridge.

\section{\protect\label{sec:Measurements-at-EMMA}Measurements at EMMA}

\begin{figure}[h]
\includegraphics[width=1\columnwidth]{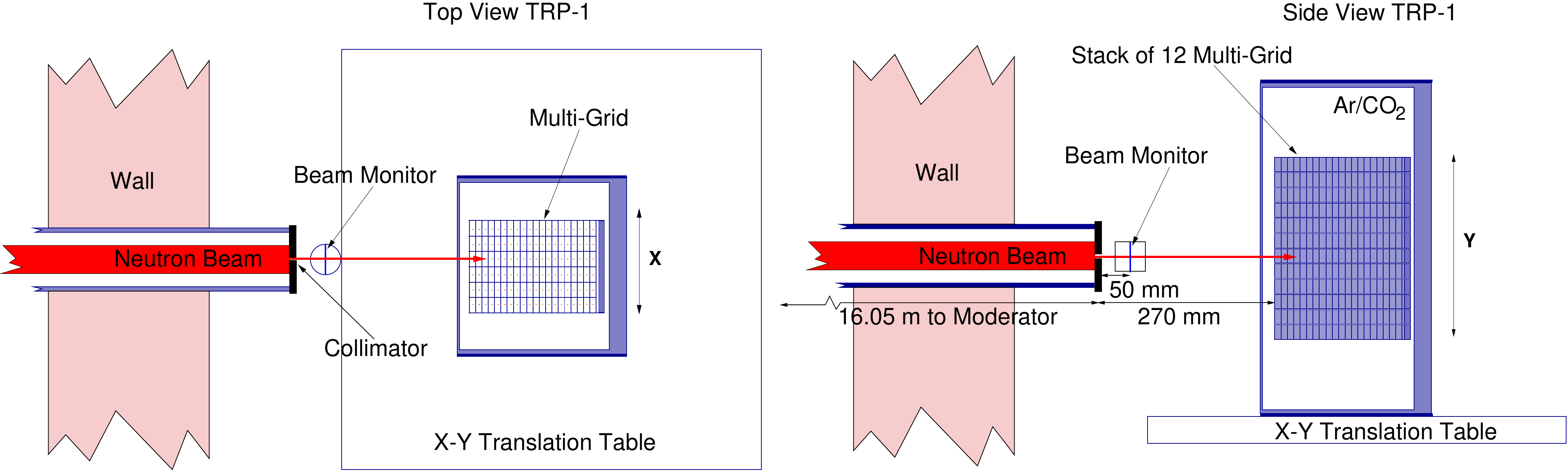}

\caption{\protect\label{fig:Schematic-of-setup}Schematic of the measurement
at EMMA, showing the positioning of the TRP-1 multi-grid prototype
with respect to the neutron beam line.}
\end{figure}

The TRP-1 prototype was tested at the EMMA beam line of the ISIS spallation
neutron source in the UK. Fig.~\ref{fig:Schematic-of-setup} displays
a simplified schematic of the test setup. The detector was mounted
on a $x-y$ translation table and aligned by scanning the direct beam
across several voxels to determine the positions where it crossed
from one voxel to the next. Beam size was determined by $\mathrm{B_{4}C}$
slits which were set to give beam dimensions of 2.0$\times0.5\:\mathrm{mm^{2}}$
in the x and y directions. Direct beam measurements employed a `white'
neutron beam which peaked in intensity at $\sim1\:\textrm{\AA}$ and
extended up to $\sim4.7\:\textrm{\AA}$. As TOF was recorded, particular
ranges of wavelength could be selected in the offline data analysis.
Accumulated neutron intensity during a run was determined from the
accumulated proton-beam charge on the spallation target and recorded
counts on the 0.25~mm-thick GS1 Li-Glass scintillator Beam Monitor
(BM) \citep{BeamMon}.

Measurements were made with the anode-wire HV ranging from 850 to
1050~V in order to determine the optimum setting. A value of 900~V
ensured that the VPC operated well within the proportional-mode voltage
range and all electronic components operated well within their linear
dynamic ranges. All measured and simulated distributions shown in
this article are at an HV of 900~V. Both anode and cathode signals
are necessary to determine which voxel has fired. However for brevity
only anode-wire distributions are shown, as those from the cathode
grids are similar.

To ensure efficient rejection of gamma-ray events, the gamma response
of the detector was calibrated by inserting a 2~mm-thick sheet of
Cd on the front face of the detector. Cd has an extremely high neutron
capture cross section, so that it blocks neutrons, at the same time
producing gamma rays from the radiative capture process. These convert
to electrons in the materials of the MG detector, mainly by Compton
scattering, which produce a small signal in the VPC. The resulting
enhanced-gamma, PH spectra were then used to set the discriminator
thresholds to minimise the gamma sensitivity of the detector \citep{Gamma-calib}.

\begin{figure}[h]
\includegraphics[width=1\columnwidth]{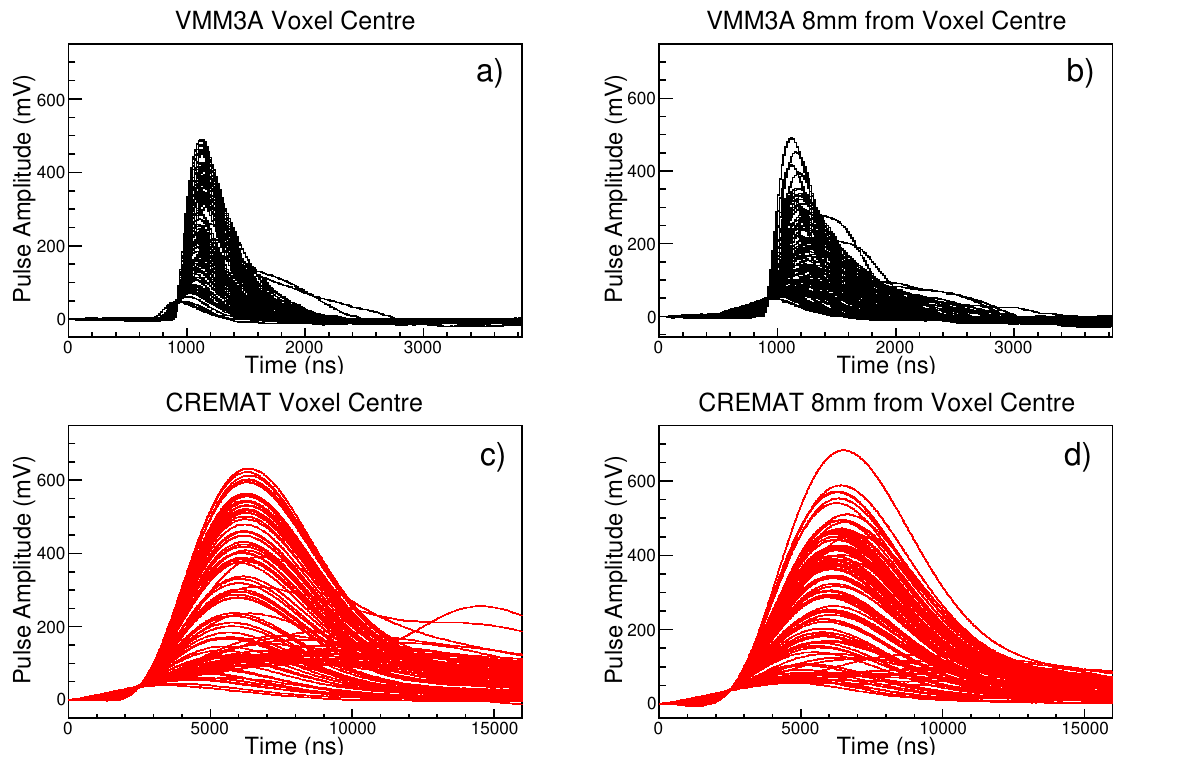}

\caption{\protect\label{fig:Waveforms-from-VMMA}Sample of 100 anode-wire waveforms
from VMMA and CREMAT electronics, taken at voxel $x$ coordinates
0.0 and 8.00~mm}
\end{figure}

The neutron beam was scanned in the x direction (horizontal) across
the face of a MG voxel. Measurements were made at x coordinates from
-13 to +13~mm (with respect to the voxel centre), in steps of 1~mm
and this was performed for TRP-1 with both VMM3A and CREMAT readout.
Samples of recorded waveforms from VMM3A and CREMAT shaping amplifiers
are displayed in Fig.~\ref{fig:Waveforms-from-VMMA}. The arrival
of a pulse triggers the digitiser, setting the time-zero point on
the recorded waveform. Thus delays caused by slow charge collection
near the voxel corners were not preserved (see Sec.~\ref{sec:Monte-Carlo-Simulations}).

Comparison of VMM3A waveforms taken at $x=0.0$~mm (Fig.~\ref{fig:Waveforms-from-VMMA}a)
and $x=8$.0~mm (Fig.~\ref{fig:Waveforms-from-VMMA}b) shows that
the former has a more sharply defined maximum PH at around 480 mV
corresponding to the maximum energy loss of an alpha produced in the
neutron capture process. The pulse shapes are also quite similar,
while at $x=8.0$~mm the pulse shapes vary considerably, frequently
with extended tails. 

With CREMAT, there is little apparent difference in pulse shape at
$x=0.0$ (Fig.~\ref{fig:Waveforms-from-VMMA}c) and 8.0~mm (Fig.~\ref{fig:Waveforms-from-VMMA}d),
while there does appear to be a slightly sharper cut off at maximum
PH of around 600 mV for $x=0.0$~mm. 

\begin{figure}[h]
\includegraphics[width=1\columnwidth]{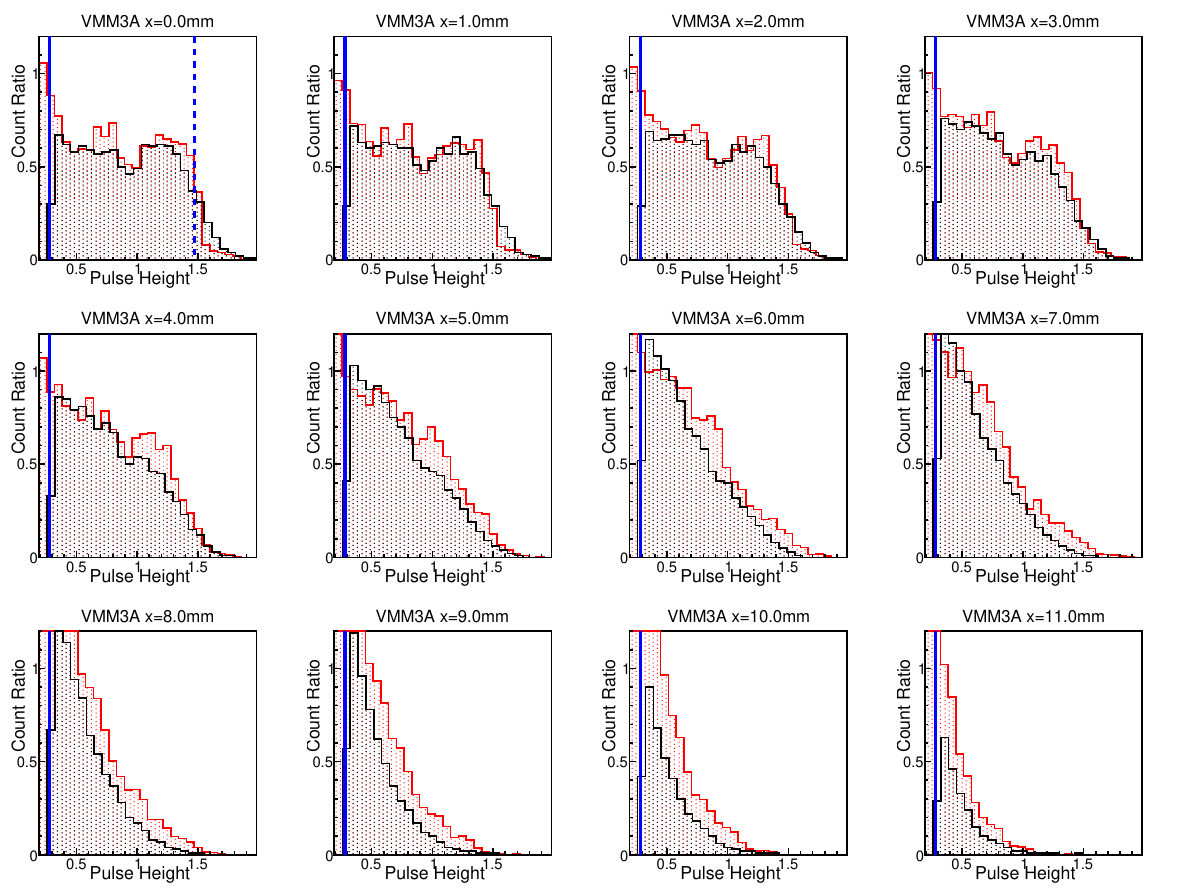}

\caption{\protect\label{fig:Pulse-height-VMM3A}PH spectra obtained for the
position scan with VMM3A readout. Pulse-height alignment is described
in the text. Count Ratio is the quotient of TRP-1 and BM counting
rates. 12 measured spectra (black shading) are displayed for positions
x=0.0 (top left plot) to x=11.0~mm (bottom right) in 1~mm steps.
Equivalent Garfield++ simulations are red shaded.}
\end{figure}

PH spectra obtained with the VMM3A system are displayed in Fig.~\ref{fig:Pulse-height-VMM3A},
where the black-shaded distributions are measurements made with the
anode-wire potential set to 900~V. These are compared to the red-shaded,
simulated distributions, described in Sec.~\ref{sec:Monte-Carlo-Simulations}.
Both measured and simulated histograms display the summed counts from
the front six MG voxels in the direct path of the beam, divided by
the integrated counts in the BM. The Li-Glass-scintillator \citep{BeamMon}
BM (Fig.~\ref{fig:Schematic-of-setup}) has a computed (Sec.\ref{sec:Monte-Carlo-Simulations})
neutron detection efficiency of $\approx0.6$\%, averaged over wavelengths
0.5 to 4.7~$\textrm{\AA},$ the range employed in the data analysis.

Where the neutron capture (and hence starting position of the $\mathrm{^{4}He}$
or $\mathrm{^{7}Li}$ track) is close to the centre of a normal blade
($\left|x\right|\leq3$ mm) the distribution displays a $\mathrm{^{4}He}$
`bump' at the upper end of the spectrum, which washes out progressively
as the capture event moves towards the corner of a voxel. As charge
collection becomes slower and more dispersed in time, the short shaping
time employed by the VMM3A integrates less of the total charge collected,
leading to a loss of PH and PH resolution. 

\begin{figure}[h]
\includegraphics[width=1\columnwidth]{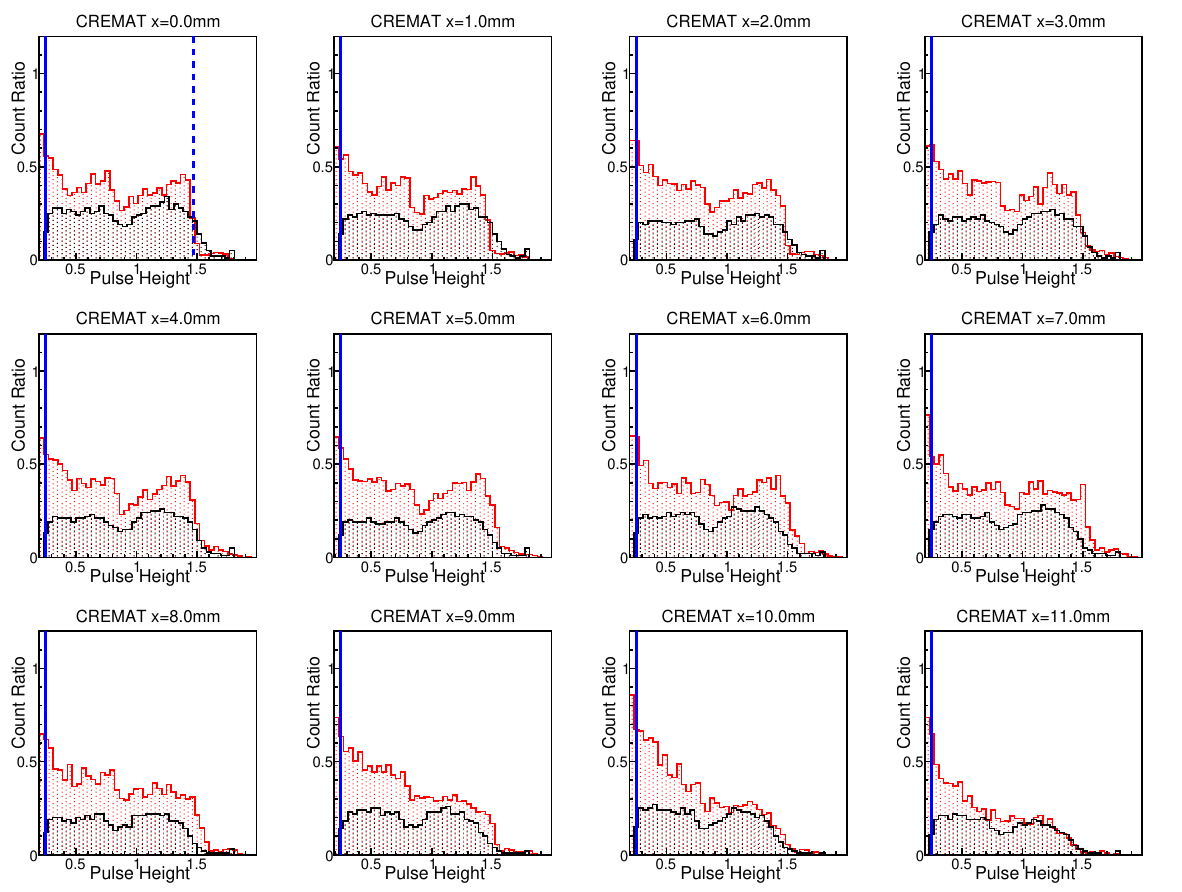}

\caption{\protect\label{fig:Pulse-height-CREMAT}PH spectra obtained for the
position scan with CREMAT readout. Axes as in Fig.~\ref{fig:Pulse-height-VMM3A}.
12 measured spectra (black shading) are displayed for positions x=0.0
(top left plot) to x=11.0~mm (bottom right) in 1~mm steps. Equivalent
Garfield++ simulations are red shaded.}
\end{figure}

Equivalent PH spectra, taken with the CREMAT readout are shown in
Fig.~\ref{fig:Pulse-height-CREMAT} (black-shaded distributions)
and compared to simulated red-shaded distributions. Counts were normalised
to the BM in the same manner as VMM3A (Fig.~\ref{fig:Pulse-height-VMM3A}).
The $\mathrm{^{4}He}$ bump at the high end of the distribution is
visible on all spectra, but with a slight decrease in counts towards
the voxel corner. The CREMAT amplifier, with a pulse peaking time
of $\approx4\:\mu s$ (a factor 20 larger than the VMM3A), integrates
the dispersed charge signals close to a corner more completely.

Since the PH conversion gains differed for VMM3A, CREMAT and the simulation,
the PH spectra have been scaled horizontally by constant factors so
that the edges of the distributions taken at $x=0.0$~mm sit at a
value of 1.47 units, as shown by the blue dashed lines in Fig.~\ref{fig:Pulse-height-VMM3A}
and Fig.~\ref{fig:Pulse-height-CREMAT}. 1.47~MeV is the maximum
energy of the alpha particle (from the dominant neutron capture channel
$n\mathrm{+^{10}B\rightarrow^{4}He+^{7}Li}+\gamma$) when it emerges
from the $\mathrm{B_{4}C}$ coating. Thus the PH scale corresponds
approximately to MeV energy. Full blue lines in Fig.~\ref{fig:Pulse-height-VMM3A}
and \ref{fig:Pulse-height-CREMAT} denote the lower integration limit
applied to the simulated ratios (see below). This limit is equivalent
to the PH threshold applied to the measured data.

\begin{figure}[h]
\includegraphics[width=1\columnwidth]{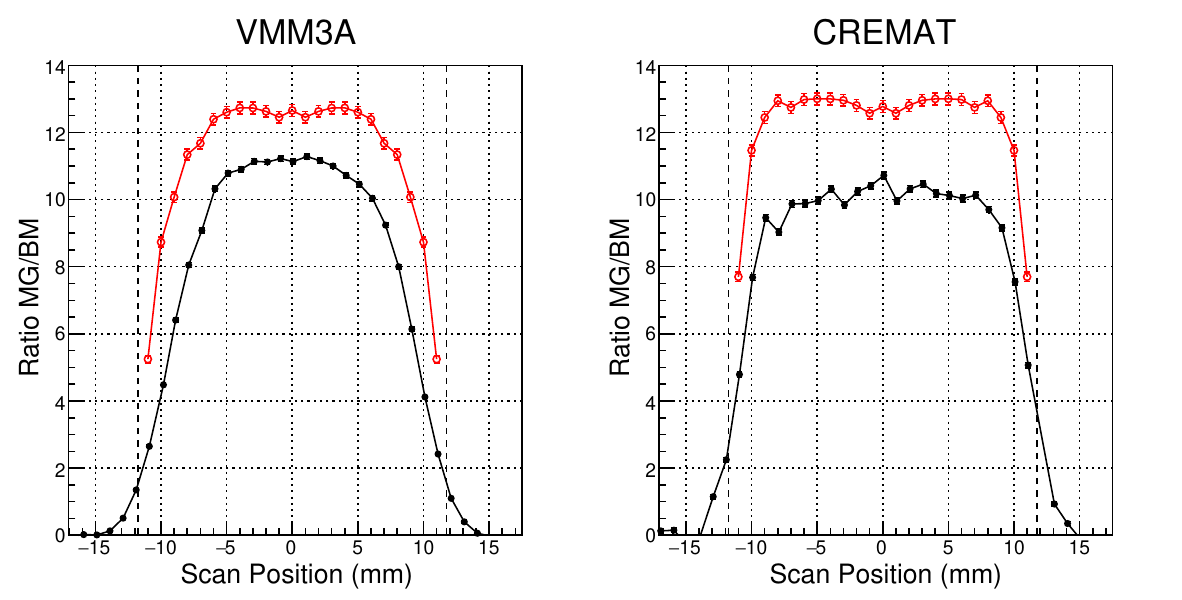}

\caption{\protect\label{fig:Relative-efficiencies}MG/BM counting ratios obtained
from the voxel x-scans. Data are shown as filled black circles, while
the open red circles show the simulations. The dashed lines denote
the edges of the voxel. Error bars, denote statistical uncertainties
and the estimated systematic error of the simulation is $\sim11$\%. }

\end{figure}

Integration of PH spectra from threshold upwards gives the ratios
of MG/BM counts shown in Fig.~\ref{fig:Relative-efficiencies}. It
is noticeable that VMM3A ratios are $\sim10$\% higher than those
from CREMAT for positions $-6\leq x\leq+6$~mm. One would expect
CREMAT, with more complete charge integration, to be better than VMM3A
and indeed the simulation predicts slightly higher values for CREMAT.
Taking the average ratios for positions $-6\leq x\leq6$~mm, the
measured value for VMM3a is 86\% of the simulated value, while for
CREMAT the measured value is 58\% of the simulated value. 

Thus it seems likely that some CREMAT signals were lost. The incident
neutron flux fluctuates due to the pulsed operation of the ISIS proton
synchrotron. If the instantaneous rate is sufficiently high, pulse
pileup and the internal data processing mechanisms of the V1470D digitiser
can lead to losses. In addition the data transfer from digitiser to
DAQ control PC has a relatively low bandwidth. VMM3A with much shorter
pulses, fast digitisation and high-bandwidth readout is much less
prone to such losses.

However the CREMAT distribution shows less loss of efficiency than
VMM3A near to the corners of a voxel and these general features are
reproduced by the simulation.

\section{\protect\label{sec:Monte-Carlo-Simulations}Monte Carlo Simulations}

A model of the PH response of a MG voxel \citep{MG_G4} has been developed
using the Garfield++ toolkit \citep{Garfield} in conjunction with
Geant4 \citep{G4_0}. $\mathrm{^{4}He}$ or $\mathrm{^{7}Li}$ products
of neutron capture on $\mathrm{^{10}B}$, are emitted from the $\mathrm{B_{4}C}$
film coating a normal blade. As they are produced back-to-back only
one of the nuclei can reach the chamber gas, while the other stops
in the normal blade. The energy of the particle, on reaching the gas,
depends on the thickness of $\mathrm{B_{4}C}$ traversed \citep{Eloss},
so that the energy and direction were sampled randomly from two-dimensional
energy-vs.-angle distributions calculated by a Geant4 model of TRP-1. 

The Geant4 model records event-by-event all neutron and gamma interactions
in TRP-1 and the BM. The produced ROOT \citep{ROOT} TTree file is
then processed to find the energy and direction of the $\mathrm{^{4}He}$
or $\mathrm{^{7}Li}$ nuclei emitted from the $\mathrm{B_{4}C}$ film
on a normal blade. Energy and polar angle are stored in a two-dimensional
ROOT histogram which the Garfield++ model uses to generate randomised
starting energy and angle. The present calculations assume $\mathrm{1\:\mu m}$
film thickness for the front voxels of TRP-1, used in the beam scan.
The thickness affects the energy loss of particles in the film before
they emerge into the chamber gas and the width of the $\mathrm{^{4}He}$
`peak' at the upper end of the PH distribution. Analysis of the
TTree file also records the number of events where a nucleus reaches
the chamber gas in the front six in-beam voxels and the integrated
counts in the Li-Glass BM. The ratio of these at $x=0$~mm is a base
value, which is then scaled by the position dependent factors calculated
by Garfield++.

In the Garfield++ model the nuclei plough through the detector gas
(80/20\% $\mathrm{Ar/CO_{2}}$ mix at STP) producing ionisation and
usually they stop in the gas. Energy loss and angle straggling were
calculated by SRIM \citep{SRIM} and read in as a look-up table by
class TrackSrim of Garfield++. The drift of charge carriers (electrons
and ions) induced by the electric field was calculated by Runge-Kutta-Fehlberg
integration as implemented by class DriftLineRKF of Garfield++. The
accumulated charge as a function of time Q(t), was calculated for
electrons and ions on both anode wire and cathode grid.

The Q(t) distributions were then post processed to convolute a `transfer
function' ($\mathcal{F}(t))$, to produce a voltage distribution
V(t) which can be compared to experiment. A single set of $10^{4}$
$\mathrm{^{4}He}$ and $\mathrm{10^{4}\:^{7}Li}$ charge accumulation
distributions was calculated for each x value of a position scan and
then post processed with either VMM3A or CREMAT pulse shape distributions
described by the relation:

\begin{equation}
\mathcal{F}(t)=g\exp(n)(\frac{t}{t_{p}})^{n}\exp(-t/\tau)\label{eq:trans}
\end{equation}

For VMM3A parameter $n=3$ was the number of amplifying stages, $\tau=67\,$ns
was the shaping time constant, $t_{p}=n\tau=200$~ns was the peaking
time and $g=-3.0$~mV/fC was the gain factor, while for CREMAT the
parameters were $n=2$, $\tau=2\:\mu$s, $t_{p}=4\:\mu$s and $g=-2.3$~mV/fC.

\begin{figure}[h]
\includegraphics[width=1\columnwidth]{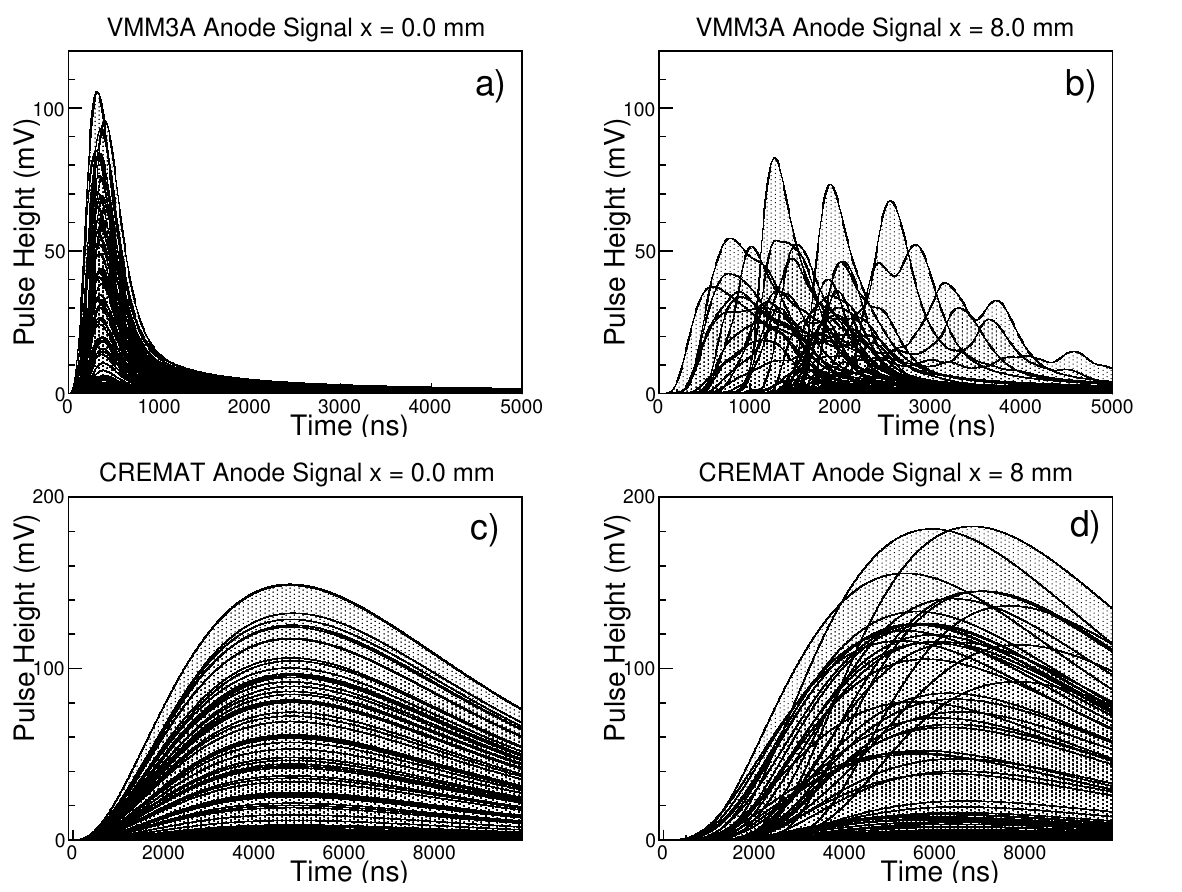}

\caption{\protect\label{fig:Simulated-pulses}The Garfield++ simulation of
pulse forms from VMM3A and CREMAT shaping amplifiers.}
\end{figure}

Fig.~\ref{fig:Simulated-pulses} shows samples of 150 simulated pulses,
calculated with VMM3A pulse shaping parameters at $x=0.0$~mm (Fig.~\ref{fig:Simulated-pulses}a),
$x=8.0$~mm (Fig.~\ref{fig:Simulated-pulses}b) and CREMAT pulse
shaping parameters at $x=0.0$~mm (Fig.~\ref{fig:Simulated-pulses}c),
$x=8.0$~mm (Fig.~\ref{fig:Simulated-pulses}d). Unlike the experimental
situation, time zero $t_{0}$ is the point at which the post-capture
nucleus emerges from the $\mathrm{B_{4}C}$ layer on a normal MG blade.
At $x=0.0$~mm pulses start very shortly after $t_{0}$ and are uniform
in shape, while at $x=8.0$~mm arrival times are spread over several
$\mu$s and shapes vary considerably. Both VMM3A and CREMAT simulated
waveform amplitudes are 20-25\% of the measured waveform amplitudes
(Fig.~\ref{fig:Waveforms-from-VMMA}), which suggests that the gas
gain calculation is too low, but this does not have a significant
impact on the relative position dependence of charge collection.

\begin{figure}[h]
\includegraphics[width=1\columnwidth]{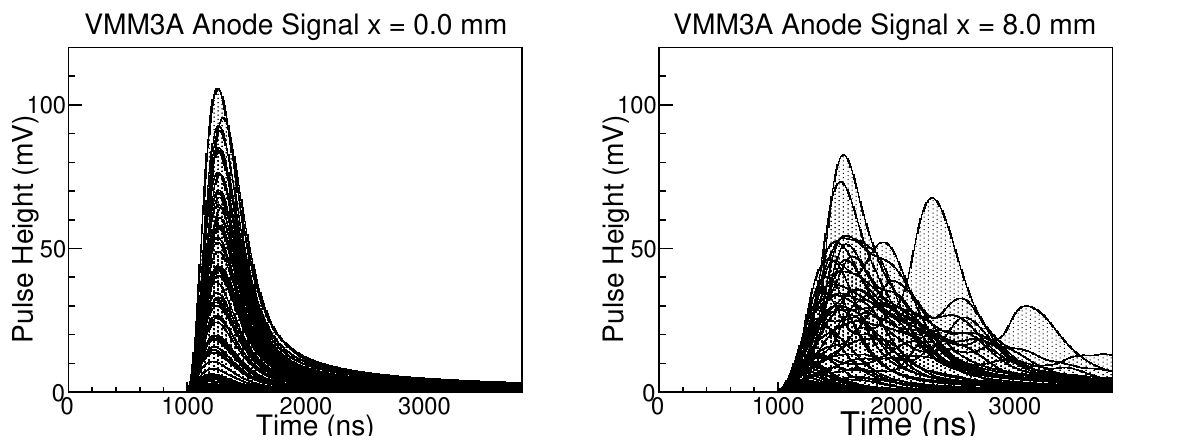}

\caption{\protect\label{fig:Shifted-simulated-pulses}The shifted simulation
of pulse forms from the VMM3A shaping amplifer.}

\end{figure}

Fig.~\ref{fig:Shifted-simulated-pulses} shows the VMM3A simulations,
where the pulses have been shifted in time to start at time $t=1000$~ns.
This mimics the experimental situation (Fig.~\ref{fig:Waveforms-from-VMMA})
where the arrival of a pulse triggered the start of the digitisation
sequence by the flash ADC. The VMM3A simulations give more of a tail
on the pulse than the observation, most visible at $x=0.0$~mm, where
the pulse shape is most uniform. \textcolor{black}{However the VMM3A
circuitry does implement ion tail cancellation and this feature is
not included in the transfer function convolution (Eq.~\ref{eq:trans}).}

For CREMAT charge collection extended up to 10~$\mu$s, while the
measurements extend to 16 $\mu$s. Nevertheless the simulation reproduces
the pulse shape up to 10~$\mu$s reasonably well.

As shown in Fig.~\ref{fig:Relative-efficiencies} the simulated MG/BM
counting ratios are larger than the measured values. They were calculated
for a $\mathrm{B_{4}C}$ thickness of 1.0~$\mu$m applied to the
normal blades of the MG. The maximum uncertainty in thickness is $\sim0.15\:\mu$m
and at a value of 0.85~$\mu$m the simulated average central VMM3A
efficiency ratio drops from 12.6 to 12.1. This leads to a systematic
uncertainty of $\approx4$\%. The ratio also depends on the computed
BM efficiency. The BM had a thickness of 0.25~mm and the composition
was as specified for glass type GS1 in Ref.\citep{BeamMon}. A threshold
of 2 MeV was applied to the simulated PH spectrum to reject gamma-ray
events. On the basis of uncertainty in the effective thickness of
$\mathrm{^{6}Li}$ in the scintillator, the systematic uncertainty
in the computed efficiency ratio is estimated to be $\sim10$\%. Added
in quadrature to the MG $\mathrm{B_{4}C}$-thickness uncertainty gives
a total uncertainty for the ratio of $\pm11$\%.

\section{\protect\label{sec:Summary-and-Conclusions.}Summary and Conclusions.}

The cuboidal voxels of the MG VPC employed for the T-REX neutron TOF
spectrometer at the ESS have low electric field gradients close to
the corners. This leads to slow charge collection in these regions,
which can reduce the pulse height of the signal and the efficiency
for detection of the charged ions produced by neutron capture on $\mathrm{^{10}B}$.
The position dependence of the pulse-height response of a T-REX prototype,
TRP-1, has been measured using a finely collimated thermal neutron
beam at the EMMA test beam line of ISIS. The beam was scanned across
a small group of voxels at the front of TRP-1 and the charge collected
from the anode wires and cathode grids of the VPC was read out by
two alternative sets of electronics based on CREMAT or VMM3A technology.

The major determinant of the pulse-height uniformity across a position
scan was the time constant of the shaping amplifier, which effectively
sets the integration period of charge collection. A ratio, the quotient
of TRP-1 and beam monitor counting rates, was measured. CREMAT, with
a pulse peaking time of $4\:\mu s$ showed better pulse-height uniformity
than VMM3A with a pulse peaking time of 200~ns and this led to less
reduction in the ratio at positions close to voxel corners. However
VMM3A showed a higher ratio closer to the centre of a voxel. 

The pulse-height response and efficiency of TRP-1 have been modelled
using a combination of Geant4 and Garfield++. Geant4 gave the probability
of neutron capture and the energy distributions of post-capture ions
entering the VPC gas, while Garfield++ calculated the time dependence
of the charge collected at the anodes and cathode. The model predicts
the measured position dependence of pulse-height response quite well
for both VMM3A and CREMAT. For VMM3A the measured MG/BM counting ratio
was 86\% of the predicted value, which has an estimated systematic
uncertainty of 11\%. However for CREMAT the measurement was only 58\%
of the predicted value. 

Thus there was a significant loss of counting efficiency with CREMAT
readout which produced long pulses more prone to pileup effects and
also required an external digitiser, which was read out via a fairly
low bandwidth link. Pileup was expected to be minimal with the VMM3A,
which produced pulses a factor 20 shorter than CREMAT. VMM3A also
has fast onboard digitisation and high bandwidth links to output the
data, so that counting efficiency losses were expected to be small.
In this respect it is ideal for a large spectrometer operating at
high rates. The 200 ns peaking time of the VMM3A shaping amplifier
is short of optimum and the present Garfield++ simulations predict
that a modest increase to 500~ns would bring the position dependence
of detection efficiency very close to the $4\:\mu$s peaking-time
(CREMAT) level. This however is not technically feasible with current
hardware. With VMM3A readout, the T-REX MG will show drops in efficiency
at the horizontal boundaries between voxels, but probably more confined
than the gaps in efficiency between the cylindrical $\mathrm{^{3}He}$
counters installed at several large neutron TOF arrays.

Although VMM3A pulse shaping is not optimum, its high rate capability,
fast data transfer and high channel density are ideally suited to
T-REX, operating in the high neutron flux environment at ESS. We have
concluded that the small losses in detection efficiency, resulting
from the short shaping time, are acceptable and that VMM3A-based systems
are a suitable platform for T-REX readout.

\section*{Acknowledgements}

We wish to thank the following for their invaluable assistance:
\begin{itemize}
\item the ISIS staff for the efficient provision of the neutron beam at
the EMMA facility, 
\item the technicians of the ESS Detector Group and the University of Glasgow,
engaged in the design and construction of Multi-Grid structures and
electronics.
\end{itemize}
\textcolor{black}{The University of Glasgow acknowledge that this
work has been generated in collaboration with and through financial
support by European Spallation Source ERIC under Contract 325103.
Glasgow also acknowledge support from the UK Science and Technology
Facilities Council, Grant ST/V00106X/1.}

\end{document}